\documentclass[aps,prb,amsmath,amssymb,twocolumn,showpacs,superscriptaddress]{revtex4-1}
\usepackage[pdftex]{graphicx}
\usepackage[colorlinks=true,citecolor=blue]{hyperref}
\usepackage{braket}

\newcommand{\up}{\uparrow}
\newcommand{\down}{\downarrow}
\newcommand{\tL}{\text{L}}
\newcommand{\tR}{\text{R}}
\newcommand{\tx}{\text}
\newcommand{\fig}[1]{Fig.~\ref{fig:#1}}
\renewcommand{\vec}[1]{\ensuremath{\mathbf{#1}}}
\newcommand{\gm}{\gamma}
\newcommand{\Gm}{\Gamma}
\newcommand{\sg}{\sigma}
\newcommand{\veps}{\varepsilon}
\newcommand{\Abs}[1]{\ensuremath{\left| #1 \right|}}
\newcommand{\dbraket}[1]{\ensuremath{ \langle \! \langle #1 \rangle \! \rangle}}
\newcommand{\kB} {k_\text{B}}

\begin{document}
\title{Detection of interactions via generalized factorial cumulants in systems in and out of equilibrium}

\author{Philipp Stegmann} \email{philipp.stegmann@uni-due.de}
\affiliation{Theoretische Physik, Universit\"at Duisburg-Essen and CENIDE, D-47048 Duisburg, Germany}

\author{Bj\"orn Sothmann}
\affiliation{D\'epartement de Physique Th\'eorique, Universit\'e de Gen\`eve, CH-1211 Gen\`eve 4, Switzerland} 

\author{Alfred Hucht}
\affiliation{Theoretische Physik, Universit\"at Duisburg-Essen and CENIDE, D-47048 Duisburg, Germany}

\author{J\"urgen K\"onig}
\affiliation{Theoretische Physik, Universit\"at Duisburg-Essen and CENIDE, D-47048 Duisburg, Germany}\date{\today}

\begin{abstract}
We introduce time-dependent, generalized factorial cumulants $C_s^m(t)$ of the full counting statistics of electron transfer as a tool to detect interactions in nanostructures. 
The violation of the sign criterion $(-1)^{m-1} C^m_s(t)\ge0$ for \emph{any} time $t$, order $m$, and parameter $s$ proves the presence of interactions.
For given system parameters, there is a minimal time span $t_\tx{min}$ and a minimal order $m$ to observe the violation of the sign criterion. 
We demonstrate that generalized factorial cumulants are more sensitive to interactions than ordinary ones and can detect interactions even in regimes where ordinary factorial cumulants fail. We illustrate our findings with the example of a quantum dot tunnel coupled to electronic reservoirs either in or out of equilibrium.
\end{abstract}

\maketitle

% - - - - - - - - - - - - - - - - - - - - - - - - - - - - - - - - - - - - - -
\section{Introduction}
The stochastic nature of electron transfer in mesoscopic conductors gives rise to both thermal and shot noise.~\cite{blanter_shot_2000} A deviation of the shot-noise power from the value expected for Poissonian processes of uncorrelated charge transfer indicates correlations. In particular, an enhanced, super-Poissonian shot-noise power is a clear signature of the presence of interactions.~\cite{lesovik_excess_1989,buttiker_scattering_1990} Various scenarios for super-Poissonian shot noise of currents sustained by electron tunneling have been studied both theoretically~\cite{cottet_positive_2004, engel_asymmetric_2004, novotny_shot_2004, koch_franck-Condon_2005, thielmann_cotunneling_2005, sanchez_resonance_2007, sothmann_mesoscopic_2012} and experimentally.~\cite{safonov_enhanced_2003, onac_shot-Noise_2006, barthold_enhanced_2006, chen_positive_2006, zarchin_electron_2007, zhang_noise_2007, kiesslich_noise_2007}

Full information about the counting statistics of charge transfer~\cite{levitov_electron_1996, bagrets_full_2003} is contained in the probability distribution $P_N(t)$ that $N$ charges have been transferred through the system in time $t$. From its Laplace transform, the moment-generating function ${\cal M}(z,t) := \sum_N e^{N z} P_N(t)$, one can derive moments $M^m(t):= \partial_z^m {\cal M}(z,t)|_{z=0}$ and cumulants $C^m(t):= \partial_z^m {\ln \cal M}(z,t)|_{z=0}$ as the $m$th derivative ($m\ge 1$) with respect to $z$ taken at $z=0$. The $m$th moment is the expectation value of the $m$th power, $M^m(t) = \Braket{N^m}(t)$, with $\Braket{\cdots}:=\sum_N \cdots P_N(t)$.
For at least two reasons, however, it is advantageous to study cumulants $C^m(t)=\dbraket{N^m}(t)$ instead of moments.~\cite{remark_cum} First, if the charge transport can be separated into statistically independent subprocesses, the cumulants of the total transferred charge are simply given by the sum of the cumulants of all channels. Second, in the long-time limit, all the cumulants $C^m(t)\propto t$ grow linearly in $t$ while the moments $M^m(t)\propto t^m$ grow with different powers.~\cite{bagrets_full_2003} Higher-order cumulants have been calculated for electron transport through various interacting systems.~\cite{boerlin_full_2002, pilgram_statistics_2008, cuevas_full_2003, johansson_full_2003, levitov_counting_2004, belzig_full_2005, braggio_full_2006, gogolin_towards_2006, emary_frequency-dependent_2007, flindt_counting_2008, urban_coulomb-interaction_2008, lindebaum_spin-induced_2009, schmidt_charge_2009, flindt_counting_2010, schaller_counting_2010, dominguez_electron_2010, urban_nonlinear_2010, braggio_superconducting_2011} Experimentally, the cumulants of electron tunneling through quantum dots have been measured up to 20th order by monitoring the charge occupancy of the quantum dot via a capacitively coupled quantum point contact.~\cite{gustavsson_counting_2005, fujisawa_bidirectional_2006, gustavsson_measurements_2007, fricke_bimodal_2007, flindt_universal_2009, gustavsson_electron_2009, fricke_high_2010, fricke_high-order_2010, komijani_counting_2013}

An enhancement of the shot-noise power due to interactions can be quantified by the Fano factor, i.e., the ratio of the second to the first cumulant. This raises the question whether there is additional information about interactions contained in the higher-order cumulants $m\ge 3$. Higher-order cumulants have been shown~\cite{flindt_universal_2009} to oscillate, as a universal feature, as a function of any system parameter, measurement time $t$, or order $m$, independent of whether or not interactions are present. As an alternative probe of interactions, factorial cumulants have been suggested instead.~\cite{kambly_factorial_2011, kambly_time-dependent_2013} The latter are defined as cumulants $\dbraket{N^{(m)}}$ of the factorial power $N^{(m)} := N(N-1)\cdots(N-m+1)$. For noninteracting systems, factorial cumulants do not change sign as a function of any system parameter or time $t$ and display only trivial sign changes $(-1)^{m-1}$ as a function of $m$. Any deviation from this behavior proves the presence of interactions. Interactions are, however, only a necessary but not a sufficient criterion to obtain nontrivial sign changes of factorial cumulants. In fact, the results of a recent measurement of hole transfer through an interacting quantum dot~\cite{komijani_counting_2013} could be explained within an effectively noninteracting model.

% - - - - - - - - - - - - - - - - - - - - - - - - - - - - - - - - - - - - - -
\section{Generalized factorial cumulants}
In this paper, we present a more sensitive and versatile indicator of interactions based on generalized factorial cumulants. To this end, we define the generalized factorial-moment generating function
\begin{equation}
	{\cal M}_s(z,t):=\frac{\sum\limits_{N=0}^\infty (z+s)^N P_N(t)}{\sum\limits_{N=0}^\infty s^N P_N(t)} \, ,
\end{equation}
where $N$ is the number of transferred charges, counting the charges that, say, leave the central part of the system (e.g., quantum dot) into some leads while not counting those entering; therefore, by definition, $N\ge 0$. From the generating function, we obtain generalized factorial moments $M^m_s(t):= \partial_z^m {\cal M}_s(z,t)|_{z=0}$ and cumulants $C^m_s(t):= \partial_z^m {\ln \cal M}_s(z,t)|_{z=0}$. Defining an $s$-dependent expectation value $\braket{\cdots}_s(t) := \sum_N \cdots s^N P_N(t) / \sum_N s^N P_N(t)$, one can show that $M^m_s(t) = \braket{s^{-m}N^{(m)}}_s (t)$ and $C^m_s(t) = \dbraket{s^{-m}N^{(m)}}_s (t)$, where $\dbraket{N^m}_s$ is related to $\braket{N^m}_s$ via the same recursive relation that relates $\dbraket{N^m}$ to $\braket{N^m}$.~\cite{remark_cum,remark_s=0} The factorial cumulants are recovered by setting $s=1$, i.e., $C^m_1(t)=\dbraket{N^{(m)}}(t)$.

The generalized factorial cumulants with real $s$ can be expressed in terms of the zeros $z_j(t)$ of ${\cal M}_s(z,t)$ with degeneracies $\alpha_j$ via the formula~\cite{remark_1}
\begin{equation}\label{eq:cumulantscos}
	C^m_s(t) = (-1)^{m-1} (m-1)! \sum_{j}\frac{\alpha_j \cos{[m \, \tx{arg}(-z_j)]}}{\Abs{z_j}^m}\, .
\end{equation}
The zeros are either real or appear in complex conjugated pairs. By varying $s$, their positions are simply shifted in the complex plane, i.e., $z_j(s)+s$ is independent of $s$. 

For a noninteracting fermionic system, the generating function has the form
${\cal M}_s(z,t)=(\prod _j [1-p_j+p_j(z+s)])/\sum_N s^N P_N(t)$, where $p_j$ with $0\le p_j \le 1$ is the probability for a single-particle transfer to occur.~\cite{abanov_allowed_2008, abanov_factorization_2009} Hence, all zeros $z_j = 1-1/p_j -s$ lie on the real axis with $-z_j\ge s$ and positive $\alpha_j$.
If $s\ge 0$, then $\tx{arg}(-z_j) = 0$, which fixes the sign of all generalized factorial cumulants,
\begin{equation}\label{eq:factorialsigncriterion}
	(-1)^{m-1}C^m_s(t)\ge0 \, .
\end{equation}
If $s<0$, this criterion still holds for all even orders $m$. Any violation of this behavior is a clear indication of the presence of interactions. Equation~\eqref{eq:factorialsigncriterion}, which holds for {\it any} noninteracting fermionic system constitutes the main result of our paper. For $s=1$ and $m=2$, it reduces to the well-known criterion that a super-Poissonian Fano factor, $C^2(t)/C^1(t)>1$, indicates interactions.

What are the conditions on $t$, $m$, and $s$ to observe a violation of Eq.~(\ref{eq:factorialsigncriterion}) and thus prove the presence of interactions for a given system? First, at very short times $t$, all charge-transfer processes become independent from each other. Thus, all zeros of ${\cal M}_s(z,t)$ are on the real axis and Eq.~\eqref{eq:factorialsigncriterion} holds. As a consequence, $t$ must be larger than a minimal time span $t_\tx{min}$. Second, higher-order generalized factorial cumulants are favorable since in Eq.~(\ref{eq:cumulantscos}) deviations of $\tx{arg}(-z_j)$ from $0$ are amplified by the multiplication with $m$ such that $\cos{[m \, \tx{arg}(-z_j)]}$ can become negative. Third, with decreasing $s$, the zeros $z_j(s)$ are shifted towards the positive real direction in the complex plane. Again, deviations of $\tx{arg}(-z_j)$ from $0$ become larger. Moreover, Eq.~(\ref{eq:cumulantscos}) is dominated by the zeros with the largest $\Abs{z_j(s)}^{-m}$. Thus, by varying $s$, different zeros can be brought into focus.
\begin{figure}[b]
\includegraphics[scale=0.94]{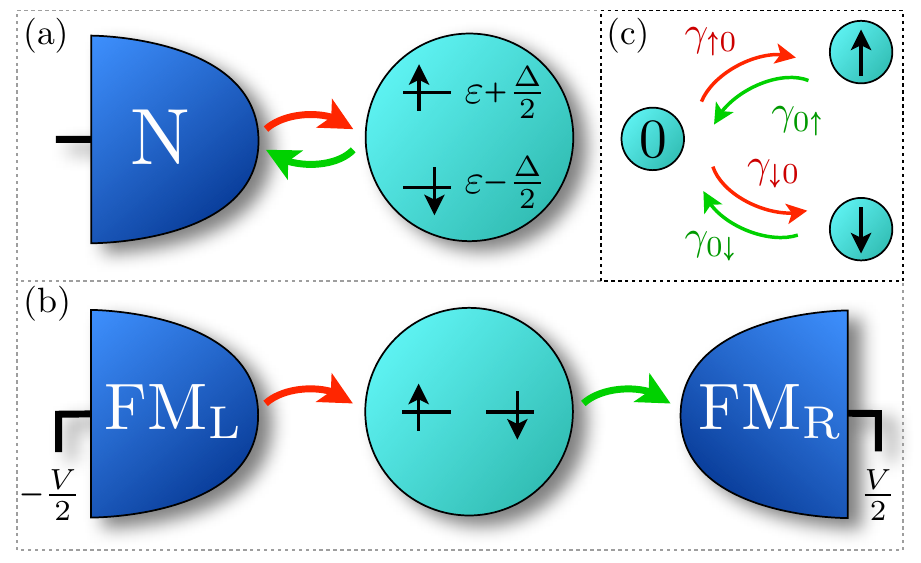}
\caption{
	(Color online) 
	(a) Equilibrium scenario: A single-level quantum dot subject to a Zeeman field is tunnel coupled to one normal lead.
	(b) Nonequilibrium scenario: A single-level quantum dot is tunnel coupled to two ferromagnetic leads with finite bias voltage.
	(c) Sketch of the states and transition rates.}
\label{fig:system}
\end{figure}
% - - - - - - - - - - - - - - - - - - - - - - - - - - - - - - - - - - - - - -
\section{Examples}
In the following, we illustrate our findings on a single-level quantum dot tunnel coupled to electronic reservoirs either in or out of equilibrium (cf. Fig.~\ref{fig:system}). A large Coulomb interaction ensures that double occupation of the dot is forbidden.
For weak tunneling (sequential tunneling), electron transfer through the dot is described by the master equation [see Fig.~\ref{fig:system}~(c)]
\begin{equation}
\begin{aligned}
\dot{P}_N^0(t)&=-\left(\gm_{\up 0} + \gm_{\down 0}\right)P_N^0(t) + \gm_{0 \up}P_{N-1}^\up(t) + \gm_{0 \down}P_{N-1}^\down(t)\, ,\\
\dot{P}_N^\up(t)&= \gm_{\up 0}P_N^0(t) -\gm_{0 \up}P_N^\up(t)\, ,\\
\dot{P}_N^\down(t)&=\gm_{\down0}P_N^0(t) -\gm_{0 \down}P_N^\down(t),
\end{aligned}
\end{equation}
for the probability $P_N^\chi(t)$ that $N$ electrons have left the quantum dot in time $t$ and the dot is in a state $\chi=0,\up,\down$ (electrons entering the dot are not counted, i.e., $N$ is {\it not} the net charge transfer between the dot and some lead). 
The transition rates $\gm_{\chi \chi'}$ from state $\chi'$ to $\chi$ are given by Fermi's golden rule. Making use of the $z$-transform $P_{z}^\chi(t) := \sum_N z^N P_N^{\chi}(t)$, we obtain
\begin{equation}\label{eq:masterZ}
	\dot{\vec{P}}_{z}(t)= \vec{W}_{z} \vec{P}_{z}(t) \, ,
\end{equation}
with the vector $\vec{P}_{z}=({P}_{z}^0,{P}_{z}^\up,{P}_{z}^\down)^T$ and the matrix
\begin{equation}
	\vec{W}_{z} =\begin{pmatrix} -\gm_{\up 0} - \gm_{\down 0}& z \gm_{0 \up} & z \gm_{0 \down}\\ \gm_{\up 0}& -\gm_{0 \up} & 0\\ 	\gm_{\down0}&0 & -\gm_{0 \down} \end{pmatrix} \, .
\end{equation}
The solution of Eq.~\eqref{eq:masterZ} is $\vec{P}_{z}(t) = \exp \left( \vec{W}_{z}t \right) \vec{P}(0)$, where $\vec{P}(0)$ is the initial probability distribution. Assuming that electron counting starts when the system has reached its steady state, $\vec{P}(0)$ is given by the stationary probability distribution, determined by $\vec{W}_1 \vec{P}(0)=0$ and $\vec{e}^T\cdot \vec{P}(0)=1$, with $\vec{e}^T=(1,1,1)$.
The moment-generating function ${\cal M}_s(z,t)= P_{z+s}(t)/P_s(t)$, with $P_{z}(t) = \vec{e}^T \cdot \vec{P}_{z}(t)$, can be expressed in terms of the three eigenvalues $\lambda_{j,z}$ ($j=1,2,3$) of the matrix $\vec{W}_{z}=\sum_j \lambda_{j,z} \vec{r}_{j,z} \otimes \vec{l}_{j,z}^{T}$ by making use of the decomposition into left and right eigenvectors $\vec l_{j,z}$ and $\vec r_{j,z}$ with normalization $\vec{l}_{j,z}^{T}\cdot \vec{r}_{j',z}=\delta_{j j'}$.
We find
\begin{equation}
	\label{eq:decomp}
	{\cal M}_s(z,t) = \sum_{j=1}^3 c_{j,{z+s}} \exp \left( \lambda_{j,z+s} t \right) / P_s(t),
\end{equation}
where we defined the amplitudes $c_{j,z}:= \left( \vec{e}^T \cdot \vec{r}_{j,z} \right) \left( \vec{l}_{j,z} \cdot \vec{P}(0) \right)$. The generalized factorial cumulants are then evaluated by performing derivatives of $\ln {\cal M}_s(z,t)$ with respect to $z$ at $z=0$.
\begin{figure}[t]
\centering
{\includegraphics[scale=0.75]{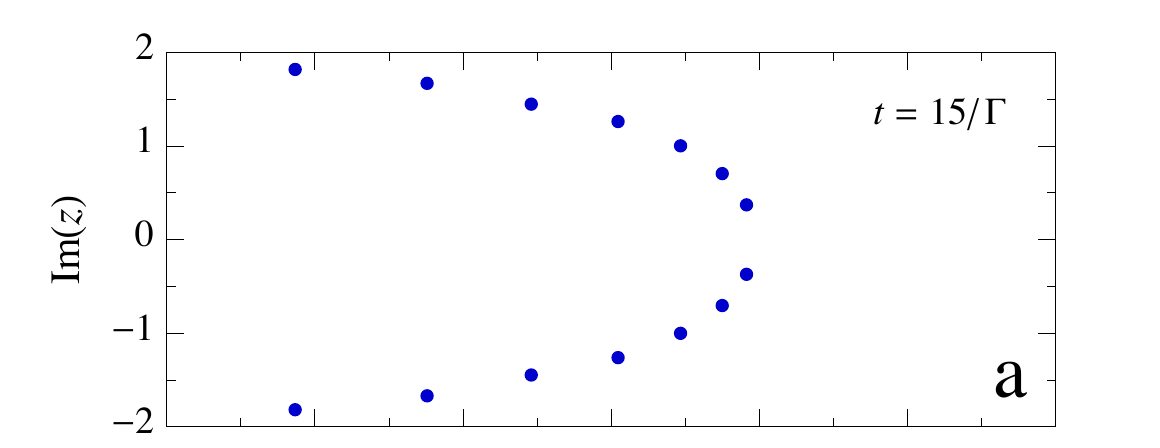}}
{\includegraphics[scale=0.75]{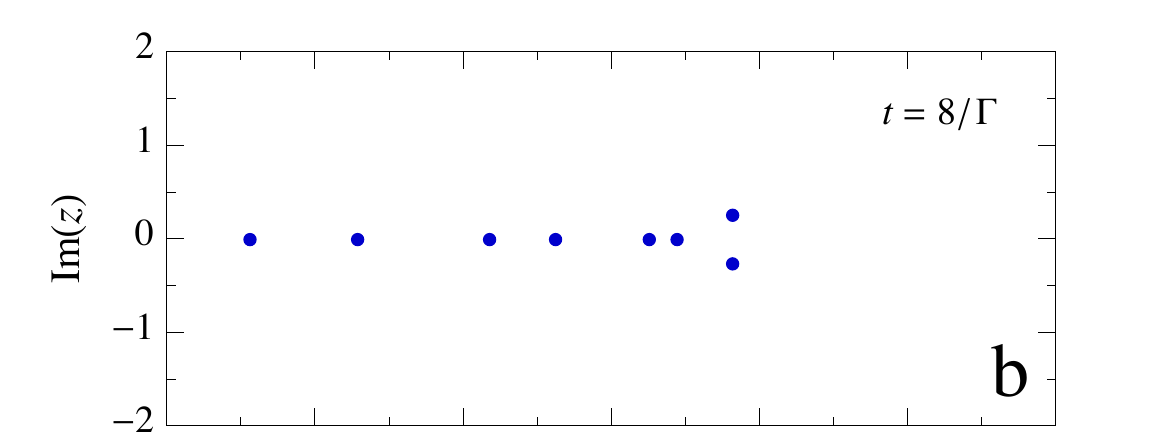}}
{\includegraphics[scale=0.75]{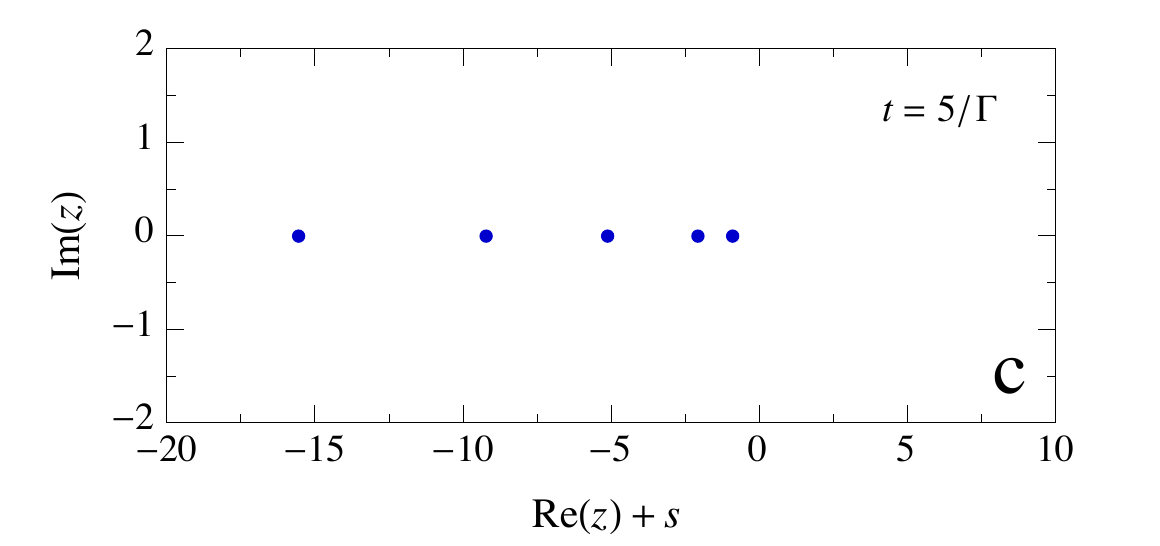}}
\caption{
	(Color online) Zeros $z_j$ for the equilibrium scenario with $\Delta=\kB T/2$, $\veps=-\Delta$, and times $t=5/\Gm, 8/\Gm, 15/\Gm$. For times $t\le t_\tx{min}\approx6.83/\Gm$, the zeros remain on the real axis. For $t>t_\tx{min}$, zeros leave the real axis and interactions can be detected by generalized factorial cumulants.
	}\label{fig:zerostmin}
\end{figure}	
\begin{figure}[t]
\centering
{\includegraphics[scale=0.75]{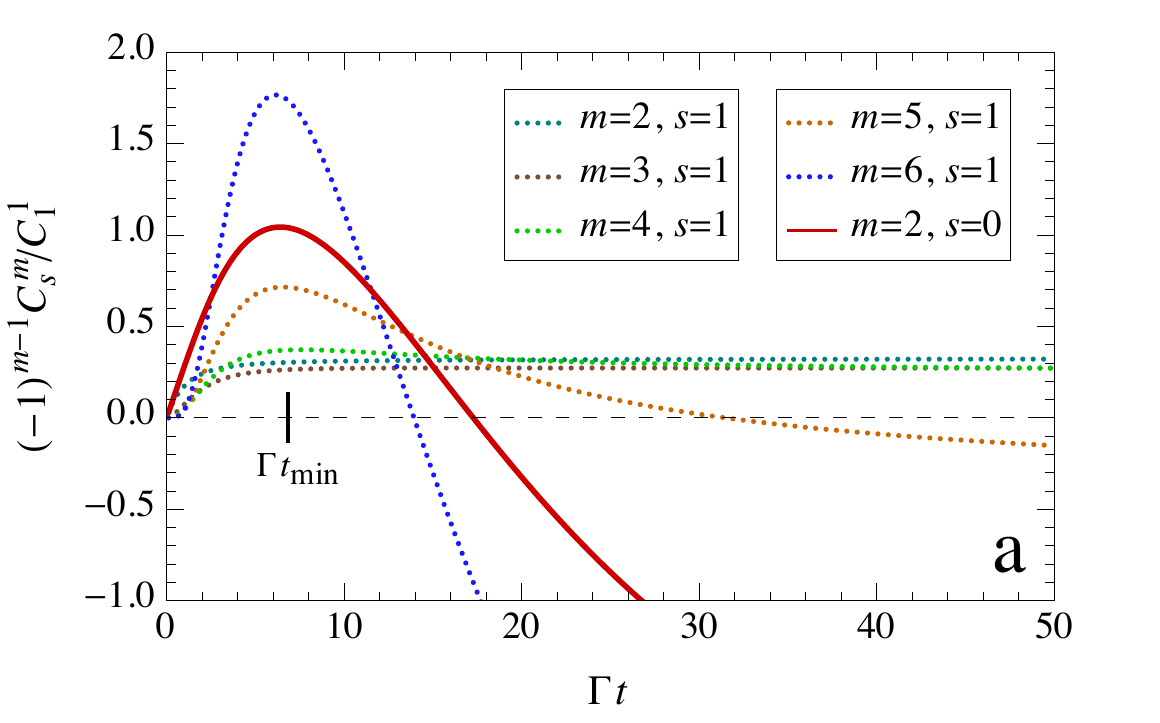}}
{\includegraphics[scale=0.75]{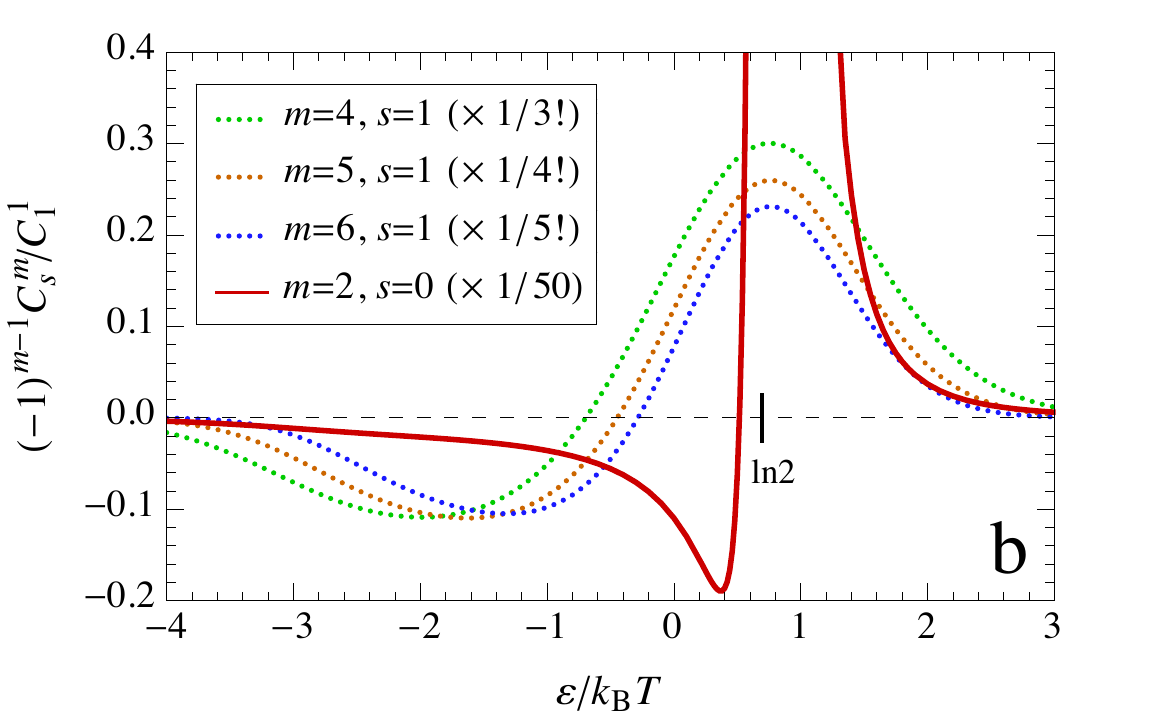}}
\caption{
	(Color online) (Generalized) factorial cumulant $C_s^m$ normalized by $C_1^1=C^1=\braket{N}>0$ for the equilibrium scenario as a function of (a) time $t$ and (b) dot-level energy $\veps$. 
	The parameters are $\Delta=\kB T/2$ and (a) $\veps=-\Delta$ or (b) $t=100/\Gm$. 
	Negative values of $(-1)^{m-1}C_s^m(t)$ indicate the presence of interactions. Interactions can be detected (a) for times larger than $t_{\tx{min}}\approx 6.83/\Gm$ and (b) for level positions $\veps \lesssim \kB T\ln 2$.
	}
\label{fig:cumteps}
\end{figure}
% - - - - - - - - - - - - - - - - - - - - - - - - - - - - - - - - - - - - - -
\subsection{Equilibrium scenario}
\begin{figure}[t]
{\includegraphics[scale=0.75]{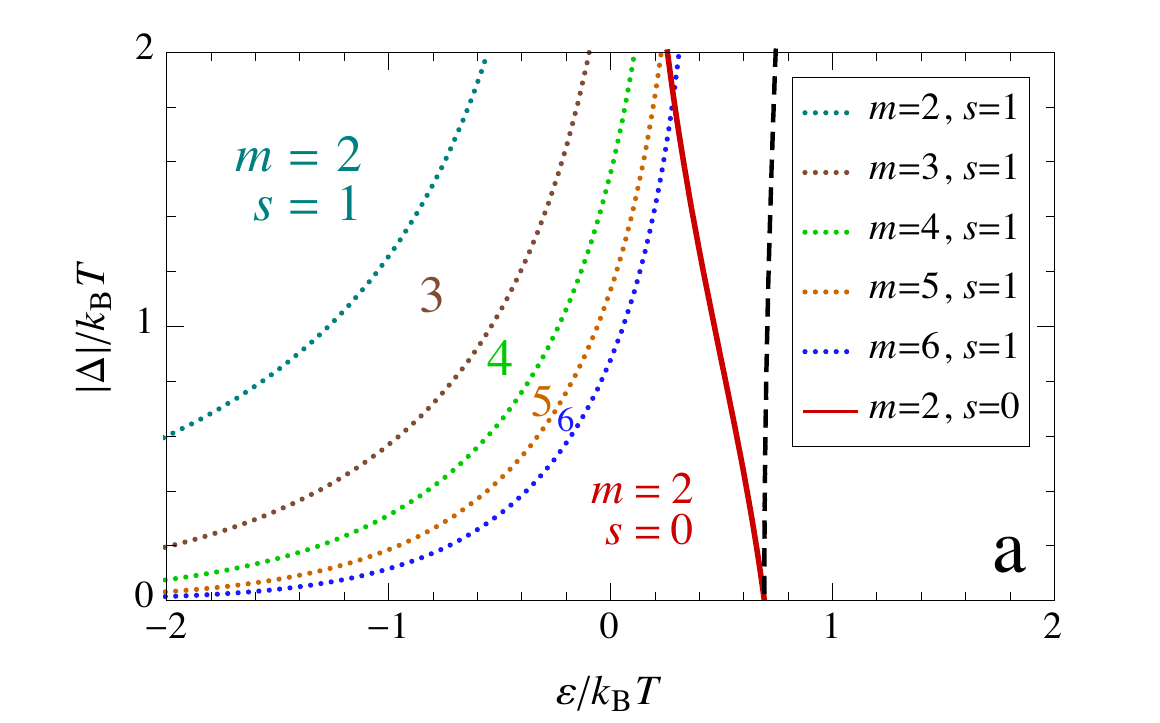}}
{\includegraphics[scale=0.75]{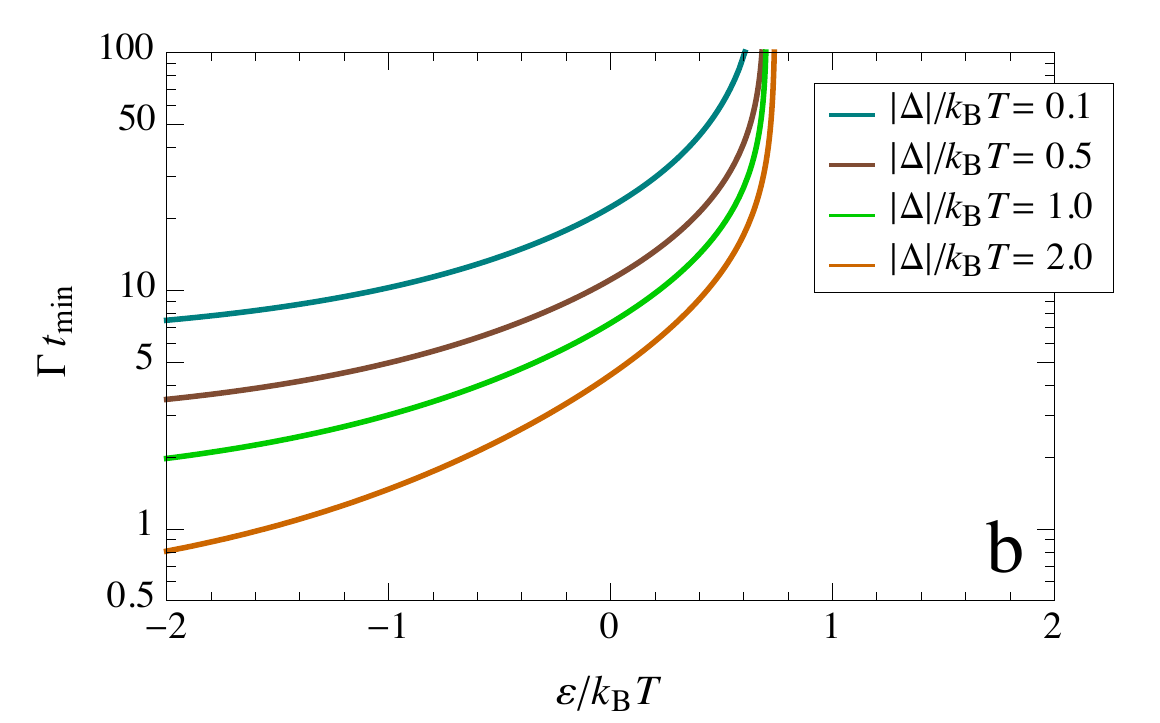}}
\caption{(Color online) (a) Parameter space and (b) minimal time span $t_\tx{min}$ of the system in equilibrium. 
	(a) To the left of the colored lines, the sign criterion for the respective $C_s^m(t)$ is violated at some time $t$. To the right of the dashed line, $\mathcal M_s(z,t)$ has only real zeros such that the sign criterion cannot be violated. (b) Minimal time span $t_\text{min}$ increases with increasing $\varepsilon$ and decreasing $|\Delta|$ and diverges at the dashed line in (a) and for $\Delta=0$.}
\label{fig:parspaceeq}
\end{figure}
\begin{figure}[t]
\centering
{\includegraphics[scale=0.75]{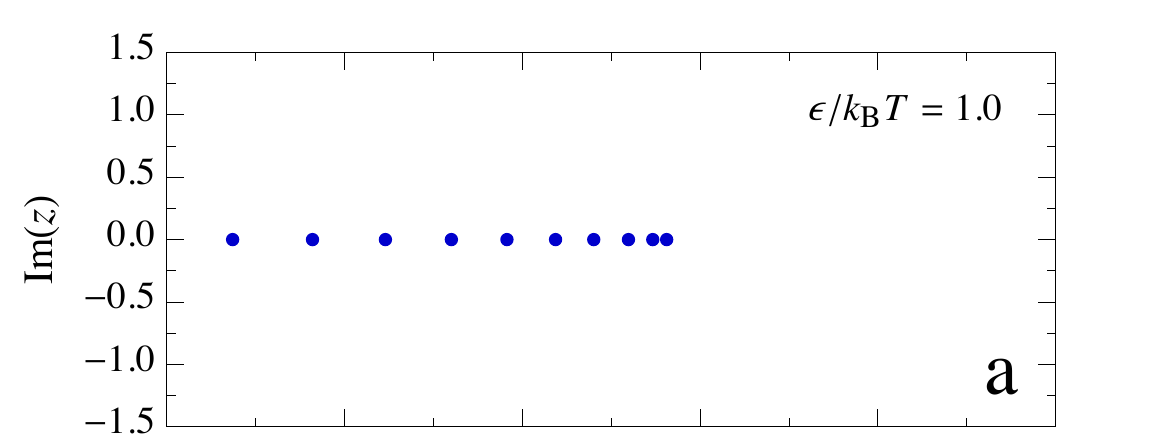}}
{\includegraphics[scale=0.75]{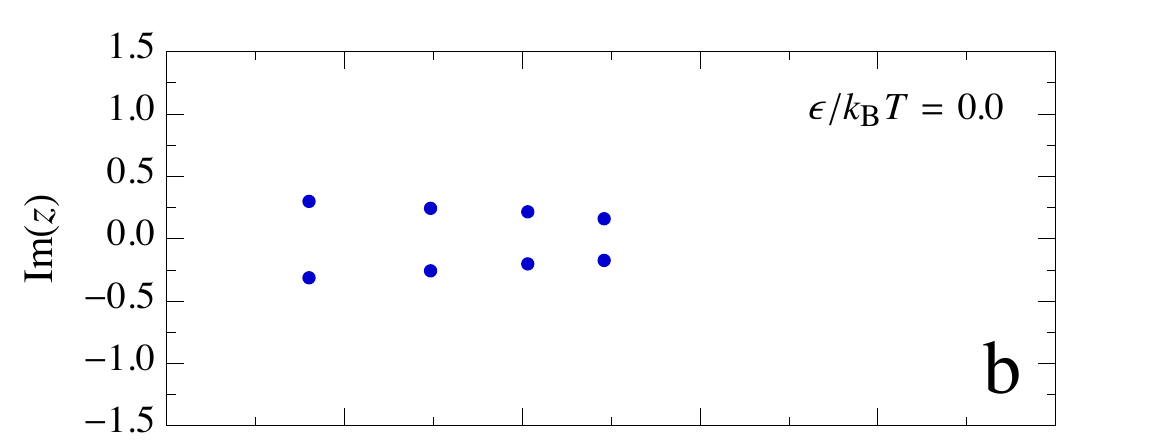}}
{\includegraphics[scale=0.75]{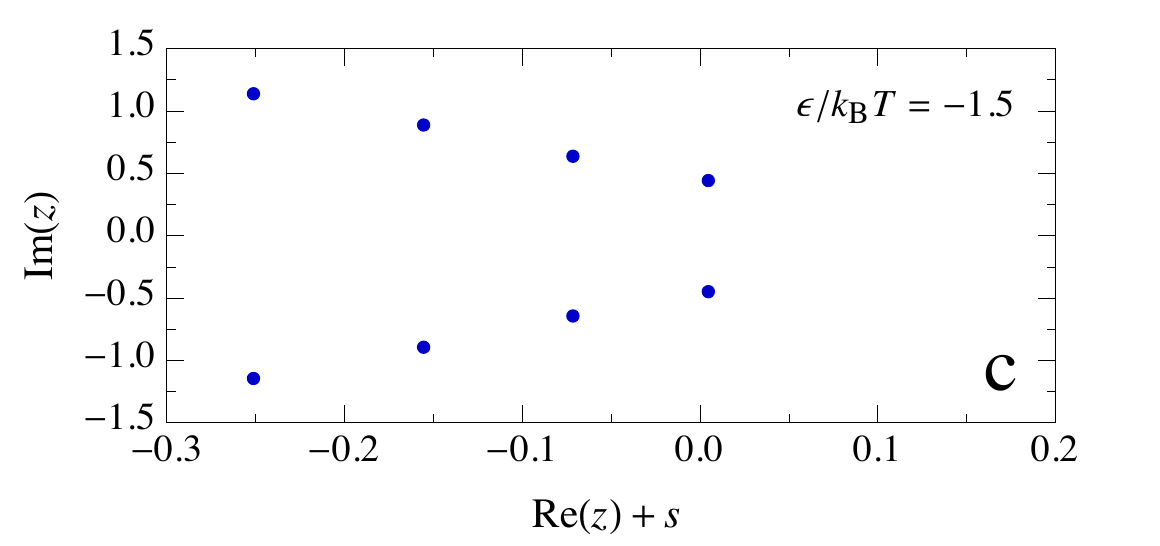}}
\caption{
	(Color online) Zeros $z_j$ for the equilibrium scenario with $\Delta=\kB T/2$, $t=100/\Gm$, and $\veps/\kB T=-1.5, 0.0, 1.0$. For $\veps/ \kB T \lesssim \ln2$, zeros leave the real axis after some time and interactions can be detected by generalized factorial cumulants.
	}\label{fig:zeroseps}
\end{figure}
As a first example, we consider an equilibrium scenario with a quantum dot coupled to one normal lead [see \fig{system}~(a)]. Spin degeneracy of the dot level is lifted by a Zeeman field, $\veps_\sg=\veps \pm \Delta/2$, where the quantum-dot level $\veps$ measured relative to the Fermi energy of the lead may be tuned by a gate voltage. The positive (negative) sign corresponds to $\sg=\up$ $(\down)$. Fermi's golden rule yields $\gm_{\sg0}=\Gm f(\veps_{\sg})$ and $\gm_{0\sg}=\Gm \left[1-f(\veps_{\sg})\right]$ with the Fermi function $f_\sg=[1+\exp(\veps_\sg/\kB T)]^{-1}$ and tunnel-coupling strength $\Gm$.
Weak tunneling corresponds to $\hbar \Gm \ll \kB T$.

For a vanishing Zeeman field, $\Delta=0$, the model can be mapped onto a noninteracting one in which a single, spinless quantum-dot level ($\chi=0,1$) is filled with rate $\gm_{10}:=\gm_{\up 0} + \gm_{\down 0}$ and emptied with rate $\gm_{01}:=\gm_{0 \up} = \gm_{0 \down}$. Only two $c_{j,z}$ are nonvanishing. As a consequence, all zeros of ${\cal M}_s(z,t)$ lie on the real axis and Eq.~(\ref{eq:factorialsigncriterion}) holds. For a finite Zeeman field, $\Delta\neq0$, the zeros of $\mathcal M_s(z,t)$ remain on the real axis for short times.~\cite{remark_short} However, after a minimal time span $t_\tx{min}$, which depends on the system parameters and can be larger or smaller than $1/\Gm$, the first pair of zeros moves from the real axis into the complex plane (cf. \fig{zerostmin}). Beyond this time, the presence of interactions can be detected from the full counting statistics, as we will detail in the following.

In \fig{cumteps}~(a), we show the factorial cumulants ($s=1$) as a function of time for fixed values of $\veps/\kB T$ and $\Delta/\kB T$. While for the first four generalized factorial cumulants Eq.~(\ref{eq:factorialsigncriterion}) holds, there is a sign change for higher orders $m$, which indicates the presence of interactions. With increasing $m$, the time at which the sign change occurs decreases and approaches $t_\tx{min}$. However, when considering generalized factorial cumulants, we observe that already $C^2_0(t)$ violates the sign criterion Eq.~\eqref{eq:factorialsigncriterion}.
In \fig{cumteps}~(b), the gate-voltage dependence of the factorial cumulants is depicted. We find that the criterion Eq.~(\ref{eq:factorialsigncriterion}) is violated for low-lying level energies $\veps$, indicating that interactions are more important in the regime when both spin states in the quantum dot have a finite occupation probability. Again, for generalized factorial cumulants with $s=0$, the possibility to detect the presence of interactions is dramatically enhanced.

Figure~\ref{fig:parspaceeq}~(a) illustrates the possibility to detect interactions via different generalized factorial cumulants. To the right of the dashed line, given by $\frac{\veps}{ \kB T}= \ln 2 + ( \frac{\Delta}{\kB T} )^2 /72 + {\mathcal O}(\frac{\Delta}{\kB T})^4$, all zeros of ${\cal M}_s(z,t)$ remain on the real axis [see \fig{zeroseps}~(a)] and $t_\tx{min}$ is infinite [see \fig{parspaceeq}~(b)]. Therefore, a violation of the sign criterion Eq.~\eqref{eq:factorialsigncriterion} and thus a detection of interactions is only possible for $\veps \lesssim \kB T\ln 2$. The second-order factorial cumulant, $s=1$ and $m=2$, violates Eq.~(\ref{eq:factorialsigncriterion}) only for rather large values of the Zeeman energy and a low-lying quantum-dot level. With increasing $m$, the region in which interactions can be detected is gradually increased to lower Zeeman splitting $\Abs{\Delta}$ and larger level positions $\veps$. For generalized factorial cumulants with $s=0$, already the second order, $m=2$, covers a much larger region of violation of Eq.~(\ref{eq:factorialsigncriterion}). This clearly demonstrates the enhanced sensitivity of generalized factorial cumulants to the presence of interactions.

% - - - - - - - - - - - - - - - - - - - - - - - - - - - - - - - - - - - - - -
\subsection{Nonequilibrium scenario}
\begin{figure}[t]
{\includegraphics[scale=0.75]{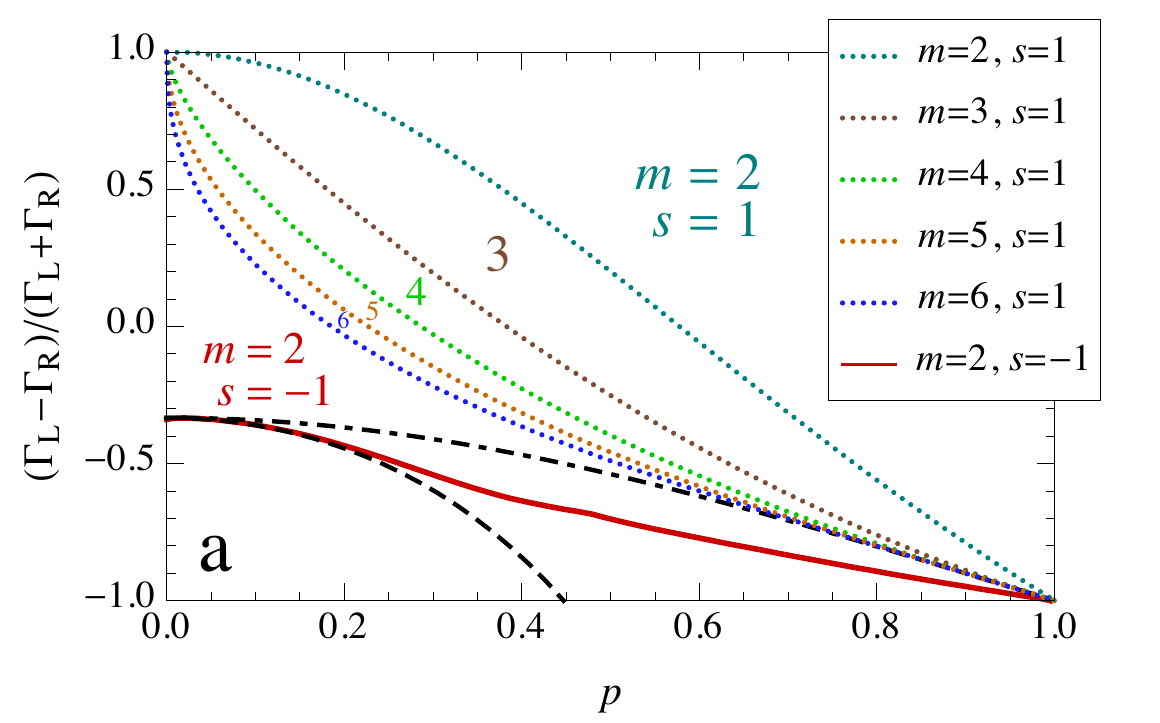}}
{\includegraphics[scale=0.75]{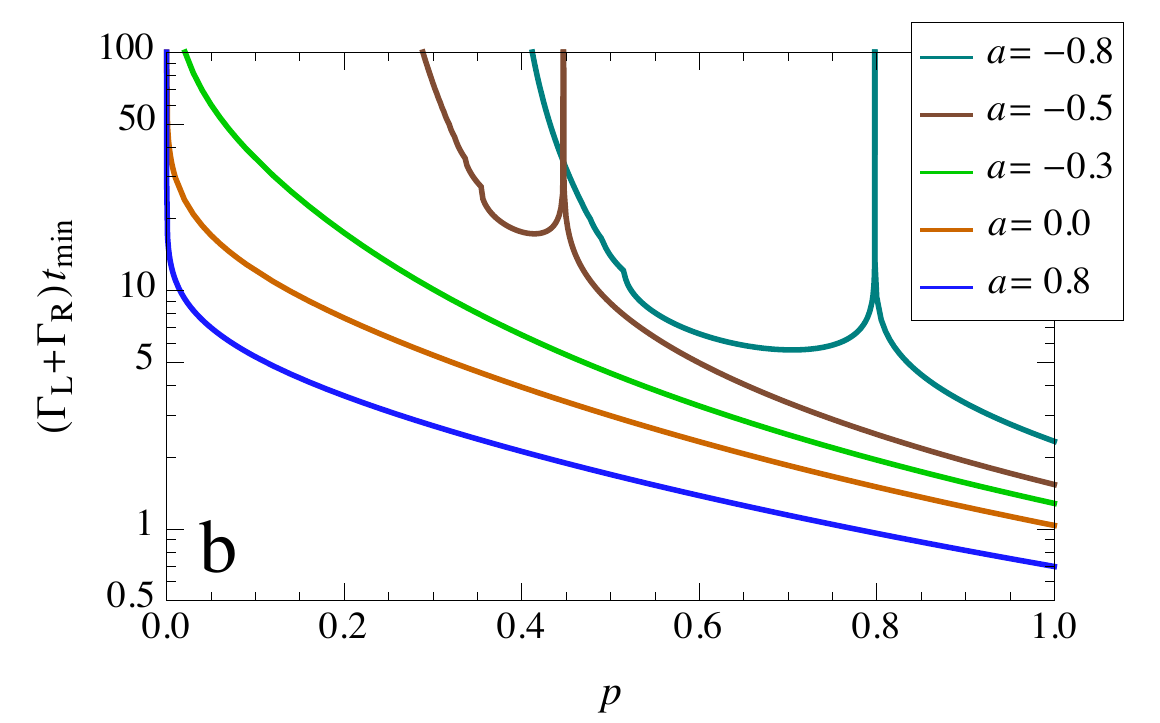}}
\caption{(Color online) (a) Parameter space and (b) minimal time span $t_\tx{min}$ of the system out of equilibrium. The polarizations are $p_\tL=p_\tR=p$, however, a different polarization $p_\tL$ (even $p_\tL=0$) does not change the results qualitatively. (a) Above the colored lines, the sign criterion for the respective $C_s^m(t)$ is violated at some time $t$. Below the dashed line, $\mathcal M_s(z,t)$ has only real zeros such that the sign criterion cannot be violated. (b) For $a\ge -1/3$, the minimal time $t_\text{min}$ increases with decreasing $p$ and decreasing $a$ and diverges at $p=0$;
for $a< -1/3$, the minimal time $t_\text{min}$ diverges both at the dashed and the dotted-dashed line in (a).}
\label{fig:parspacenoneq}
\end{figure}
\begin{figure}[t]
\centering
{\includegraphics[scale=0.75]{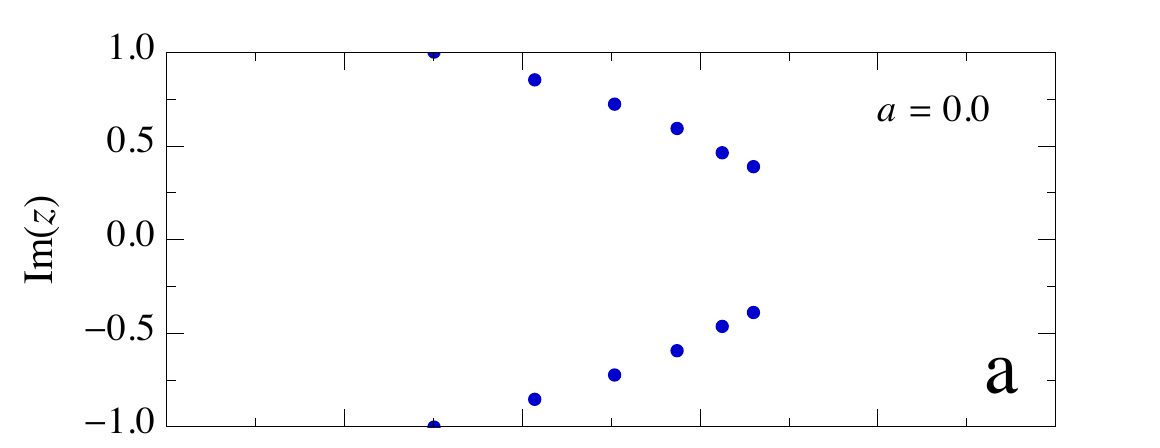}}
{\includegraphics[scale=0.75]{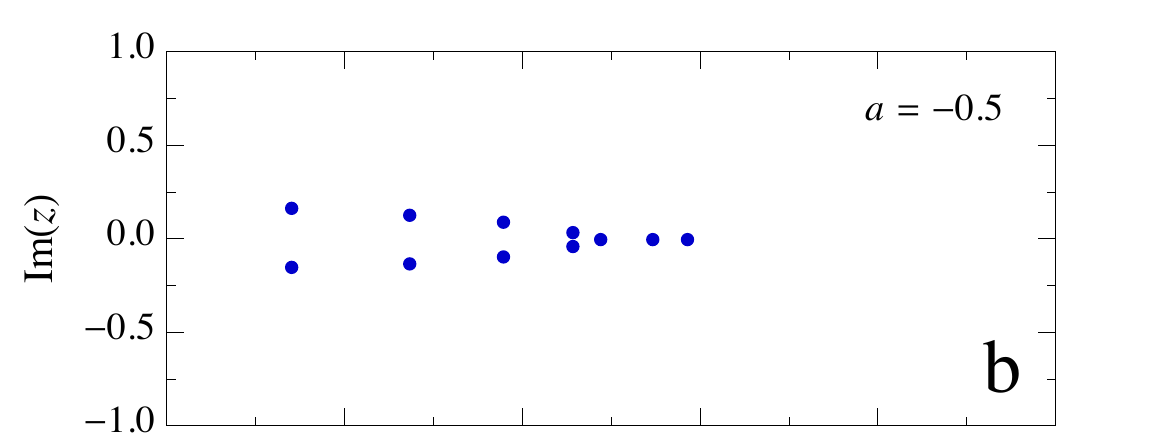}}
{\includegraphics[scale=0.75]{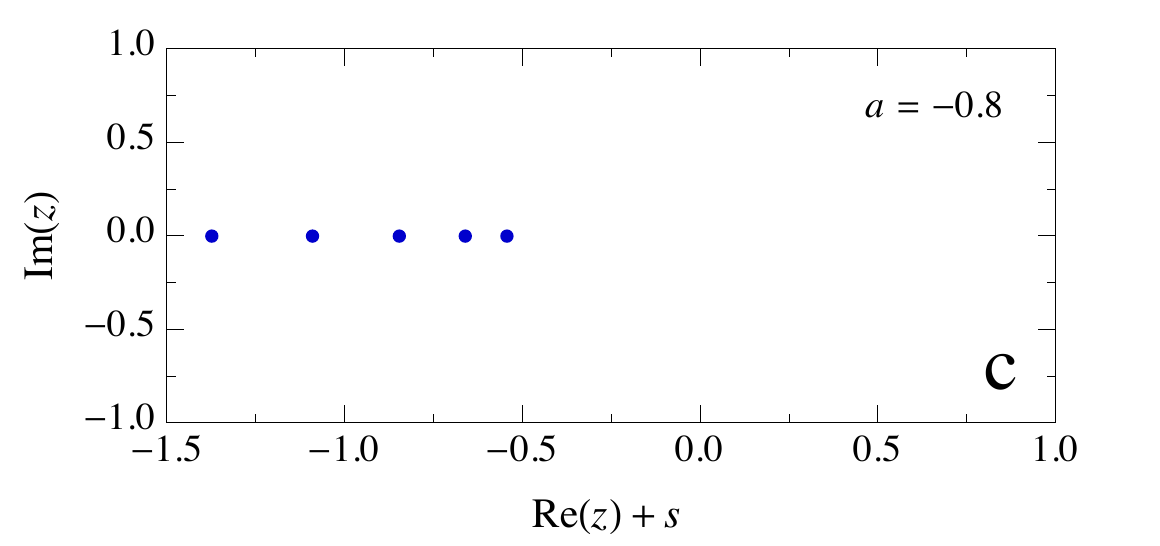}}
\caption{
	(Color online) Zeros $z_j$ for the nonequilibrium scenario discussed in \fig{parspacenoneq} with $p_\tL=p_\tR=0.35$, $t=50/(\Gm_\tL+\Gm_\tR)$, and $a=(\Gm_\tL-\Gm_\tR)/(\Gm_\tL+\Gm_\tR)=-0.8,-0.5,0.0$. For $a = -0.8$, all zeros remain on and for $a = 0.0$ all zeros leave the real axis. For $a=-0.5$, the rightmost zeros remain on the real axis, but zeros further to the left move into the complex plane. While factorial cumulants are insensitive to interactions in this case, generalized factorial cumulants can detect their presence.
	}\label{fig:zerosnoneq}
\end{figure}
As a second example [see \fig{system}~(b)], we consider a spin-degenerate quantum dot, $\Delta =0$, coupled to two ferromagnetic leads with parallel magnetizations and tunnel couplings $\Gm_r$, $r=\tx{L, R}$. A finite bias voltage $V$ applied symmetrically between the two ferromagnets gives rise to a nonequilibrium scenario. Each ferromagnet is characterized by its spin polarization $p_r$ ranging from $p_r=0$ for a normal metal to $p_r=1$ for a half-metallic ferromagnet with majority spins only. In the limit $|\veps\pm eV/2| \gg \kB T, \Gm$, transport through the quantum dot is supported by unidirectional sequential tunneling with rates 
$\gm_{\up 0}=(1+ p_\tL)\Gm_\tL$, $\gm_{\down 0}=(1- p_\tL)\Gm_\tL$, $\gm_{0 \up}=(1 + p_\tR)\Gm_\tR$, and $\gm_{0 \down}=(1 - p_\tR)\Gm_\tR$.
We denote the asymmetry of tunnel couplings to source and drain by $a=(\Gm_\tL-\Gm_\tR)/(\Gm_\tL+\Gm_\tR) $.

In Fig.~\ref{fig:parspacenoneq}~(a), we demonstrate the possibility to detect the presence of interactions in this nonequilibrium scenario. Below the dashed line, given by $a = (3 p_\tR^2+4p_\tL p_\tR+1)/(3p_\tR^2-3)$, all zeros of ${\cal M}_s(z,t)$ remain on the real axis [see \fig{zerosnoneq}~(c)], i.e., neither factorial nor generalized factorial cumulants indicate the presence of interactions.
The same is true for the trivial case $p_\tR=0$ and the dotted-dashed line, given by $a= (p_\tR^2+2p_\tL p_\tR+1)/(p_\tR^2-2p_\tL p_\tR-3)$, where the system can be mapped onto a noninteracting Hamiltonian, described by a two-state model. Of course, $t_\tx{min}$ diverges at these system parameters [see \fig{parspacenoneq}~(b)].
Above the dotted-dashed line, all the zeros move into the complex plane [see \fig{zerosnoneq}~(a)], including the rightmost zeros which dominate the behavior of the factorial cumulants, $s=1$.
For large values of the spin polarization $p=p_\tL=p_\tR$ or the asymmetry $a$ of the tunnel couplings, Eq.~(\ref{eq:factorialsigncriterion}) is already violated for the second-order factorial cumulant ($m=2$).
With decreasing $p$ or $a$, higher orders $m$ of the factorial cumulants are needed to detect interaction.

An interesting regime resides between the dashed and dotted-dashed line.
Here, the rightmost zeros remain on the real axis, but zeros further to the left move into the complex plane [see \fig{zerosnoneq}~(b)].
As a consequence, a violation of Eq.~(\ref{eq:factorialsigncriterion}) cannot be found for any ordinary ($s=1$) factorial cumulant.
This limitation is overcome by generalized factorial cumulants.
By an appropriate choice of $s$, they can probe the position of \emph{any} zero of ${\cal M}_s(z,t)$ in the complex plane and thus detect interactions where ordinary factorial cumulants fail. 
Furthermore, as in the equilibrium case, for a given order $m$, generalized factorial cumulants can detect interactions in a larger area of parameter space.

% - - - - - - - - - - - - - - - - - - - - - - - - - - - - - - - - - - - - - -
\section{Conclusions}
In summary, we proposed generalized factorial cumulants $C^m_s(t)$ of the full counting statistics of electron transport through nanostructures as a sensitive and versatile tool to detect the presence of interactions via the violation of Eq.~(\ref{eq:factorialsigncriterion}).
We found that generalized factorial cumulants are superior to ordinary ones: With decreasing $s$, interaction effects already show up in lower order $m$ and at earlier times.
This may be crucial for overcoming experimental limitations.
Furthermore, there are regimes in which general factorial cumulants can detect interactions while ordinary ones completely fail.
We illustrated our theoretical findings with two examples of a quantum dot tunnel coupled to electronic reservoirs. 
Importantly, we demonstrated that, already in a simple equilibrium case of experimental relevance, interactions can be detected via generalized factorial cumulants.
Finally, we emphasize the general validity of the criterion Eq.~(\ref{eq:factorialsigncriterion}).
It is not restricted to any specific type of interaction or transport regime and also covers multilevel and multichannel setups. It also applies to systems where coherences described by off-diagonal elements of the reduced density matrix have to be taken into account.

% - - - - - - - - - - - - - - - - - - - - - - - - - - - - - - - - - - - - - -
\acknowledgments
We acknowledge financial support from the Swiss National Science Foundation and the DFG under project KO 1987/5.
% - - - - - - - - - - - - - - - - - - - - - - - - - - - - - - - - - - - - - -
\appendix*


\begin{thebibliography}{50}
\bibitem{remark_cum}
As usual, $\dbraket{\cdots}$ is defined by the linearity property $\dbraket{\alpha N^m + \beta N^n} = \alpha \dbraket{N^m} + \beta \dbraket{N^n}$ and its relation to $\Braket{\cdots}$ via the recursive relation $\dbraket{N^m}= \Braket{N^m} - \sum_{j=1}^{m-1} \binom{m-1}{j-1} \dbraket{N^j} \Braket{N^{m-j}}$ for $m\ge 1$.
\bibitem{remark_s=0}
By shifting the index $N$ in the sum, we can rewrite $M^m_s (t) = \sum_{N=0}^\infty s^N (N+m)^{(m)} P_{N+m}(t) / \sum_{N=0}^\infty s^N P_{N}(t)$ to see that negative powers of $s$ never occur. 
Therefore, also for $s=0$, the moments and thus the cumulants are well defined, $M^m_0 (t) =m! P_m(t)/P_0(t)$.
\bibitem{remark_1}
We first limit $N$ by some $N_\tx{max}$.
Then ${\cal M}_s(z,t)=c \prod_j (z-z_j)^{\alpha_j}$ is a polynomial of order $N_\tx{max}$ and Eq.~(\ref{eq:cumulantscos}) with a finite sum is readily derived from performing the derivatives.
At the end, we let $N_\tx{max}\rightarrow \infty$.
\bibitem{remark_short}
To lowest order in $t$, Eq.~(\ref{eq:factorialsigncriterion}) takes the form $(-1)^{m-1}C^{m}_s(t)=(m-1)![\braket{N}(t)]^m+{\mathcal O}(t^{m+1})\ge 0$ for our system with $\braket{N}(t)=\frac{\gm_{0 \up} \gm_{0 \down}(\gm_{\up 0}+ \gm_{\down 0})}{\gm_{0 \up} \gm_{\down 0}+ \gm_{0 \down} (\gm_{0 \up} + \gm_{\up 0})}t$. Generalized factorial cumulants expressed by Eq.~(\ref{eq:cumulantscos}) are dominated by one zero at $-1/\braket{N}(t)$.
\end{thebibliography}
\end{document}